\documentclass[%
 reprint,
superscriptaddress,
 amsmath,amssymb,
 aps,
]{revtex4-2}

\usepackage{amsmath}
\usepackage{graphicx}
\usepackage{caption}
\usepackage{dcolumn}
\usepackage{bm}
\usepackage{siunitx}
\usepackage{wasysym}
\usepackage{xcolor}
\usepackage{subcaption}

\begin{document}

\title{Individual Nanostructures in an Epsilon-Near-Zero Material Probed with 3D-Sculpted Light}

\author{Brian Kantor}
 \affiliation{Institute of Physics, University of Graz, NAWI Graz, Universit{\"a}tsplatz 5, 8010, Graz, Austria}
 \affiliation{Christian Doppler Laboratory for Structured Matter Based Sensing, Institute of Physics, Universit{\"a}tsplatz 5, 8010, Graz, Austria}
 \affiliation{Max Planck--University of Ottawa Centre for Extreme and Quantum Photonics, 25 Templeton St., Ottawa, ON K1N 6N5, Canada}
 
\author{Lisa Ackermann}
 \affiliation{Institute of Photonic Technologies, Friedrich-Alexander-Universit{\"a}t Erlangen-N{\"u}rnberg, Konrad-Zuse-Stra{\ss}e 3/5, Erlangen, 91052, Germany}
 \affiliation{School of Advanced Optical Technologies, Friedrich-Alexander-Universit{\"a}t Erlangen-N{\"u}rnberg, Paul-Gordan-Stra{\ss}e 6, Erlangen, 91052, Germany}

\author{Victor Deinhart}
 \affiliation{Helmholtz-Zentrum Berlin f{\"u}r Materialien und Energie, Hahn-Meitner-Platz 1, 14109 Berlin, Germany}
 \affiliation{Ferdinand-Braun-Institut, Leibniz-Institut f{\"u}r H{\"o}chfrequenztechnik (FBH), Gustav-Kirchhoff-Str. 4, 12489 Berlin Germany}

\author{Katja H{\"o}flich}
 \affiliation{Helmholtz-Zentrum Berlin f{\"u}r Materialien und Energie, Hahn-Meitner-Platz 1, 14109 Berlin, Germany}
 \affiliation{Ferdinand-Braun-Institut, Leibniz-Institut f{\"u}r H{\"o}chfrequenztechnik (FBH), Gustav-Kirchhoff-Str. 4, 12489 Berlin Germany}
 
\author{Israel De Leon}
 \affiliation{Max Planck--University of Ottawa Centre for Extreme and Quantum Photonics, 25 Templeton St., Ottawa, ON K1N 6N5, Canada}
 \affiliation{ASML Netherlands B.V., De Run 6501, 5504 DR Veldhoven, The Netherlands}
 \affiliation{School of Engineering and Sciences, Tecnologico de Monterrey, Eugenio Garza Sada 2501, Monterrey 64849, NL, Mexico}
 
\author{Peter Banzer}
 \email{peter.banzer@uni-graz.at}
 \affiliation{Institute of Physics, University of Graz, NAWI Graz, Universit{\"a}tsplatz 5, 8010, Graz, Austria}
 \affiliation{Christian Doppler Laboratory for Structured Matter Based Sensing, Institute of Physics, Universit{\"a}tsplatz 5, 8010, Graz, Austria}
 \affiliation{Max Planck--University of Ottawa Centre for Extreme and Quantum Photonics, 25 Templeton St., Ottawa, ON K1N 6N5, Canada}
 \affiliation{Institute of Photonic Technologies, Friedrich-Alexander-Universit{\"a}t Erlangen-N{\"u}rnberg, Konrad-Zuse-Stra{\ss}e 3/5, Erlangen, 91052, Germany}
 \affiliation{School of Advanced Optical Technologies, Friedrich-Alexander-Universit{\"a}t Erlangen-N{\"u}rnberg, Paul-Gordan-Stra{\ss}e 6, Erlangen, 91052, Germany}
 
\date{\today}

\begin{abstract}
Epsilon-near-zero (ENZ) materials, i.e., materials with a vanishing real part of the permittivity, have become an increasingly desirable platform for exploring linear and nonlinear optical phenomena in nanophotonic and on-chip environments. ENZ materials inherently enhance electric fields for properly chosen interaction scenarios, host extreme nonlinear optical effects, and lead to other intriguing phenomena. To date, studies in the optical domain have mainly focused on nanoscopically thin films of ENZ materials and their interaction with light and other nanostructured materials. Here, we experimentally and numerically explore the optical response of individual nanostructures milled into an ENZ material. For the study, we employ 3D structured light beams, allowing us to fully control polarization-dependent field enhancements enabled by a tailored illumination and a vanishing permittivity. Our studies provide insight between complex near-fields and the ENZ regime while showcasing the polarization-dependent controllability they feature. Such effects can form the basis for experimental realizations of extremely localized polarization-controlled refractive index changes, which can ultimately enable ultrafast switching processes at the level of individual nanostructures. 
\end{abstract}

\maketitle

\section{Introduction}	

Birthed from the metamaterials community, epsilon-near-zero (ENZ) materials have been pursued over the past two decades for their intriguing optical properties \cite{alu2007epsilon,maas2013experimental}. In the linear regime, ENZ materials have already shown to enable an array of optical processes ranging from enhanced spin-orbit coupling \cite{eismann2022enhanced}, to phase modulation in photonic waveguides \cite{reines2018compact}, and tunneling through microwave cavities \cite{silveirinha2006tunneling}. Another unique behavior specific to ENZ materials lies within transparent conducting oxides (TCO), which possess a zero-crossing in their permittivity. TCOs allow for extreme refractive index changes by boosting the nonlinear Kerr response of the material due to the field enhancements promoted by the ENZ regime \cite{alam2016large}. Such nonlinear properties can be used for enhanced high-harmonic wave generation, \cite{luk2015enhanced,de2017nested,yang2019high} and are additionally being considered for purely passive phase modulators for silicon photonic technologies \cite{sha2022all,navarro2022ultrafast}. Remarkably stronger nonlinear effects and refractive index changes have been realized by combining plasmonic nanostructures with unstructured isotropic ENZ films \cite{alam2018large}.

\begin{figure*}[!t]
	\centering
	\includegraphics[width=0.75\textwidth]{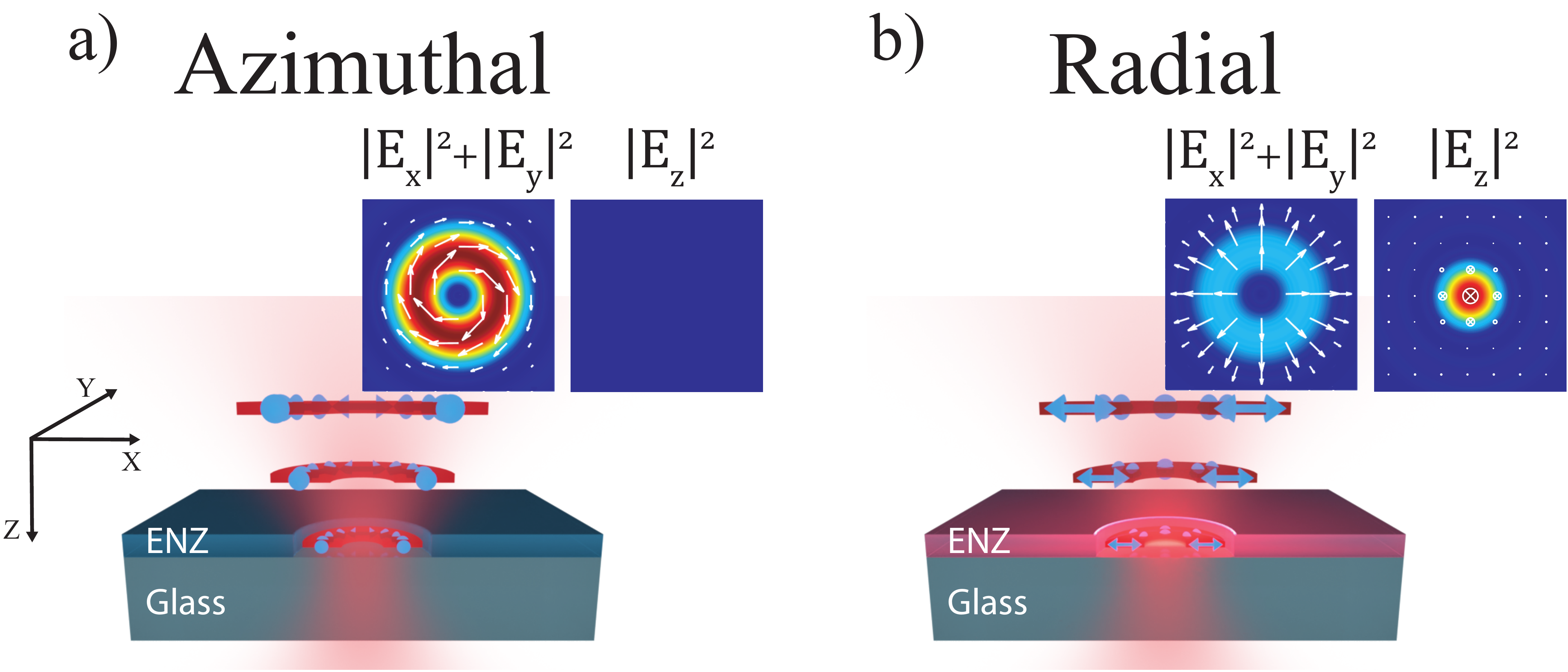}
	\caption{Sketch of a cross-sectional slice representing the two transverse field components focused onto an ideal lossless ENZ hole. The respective transverse and longitudinal field intensity distributions at the focus (in vacuum) are also shown as insets with field vectors as an overlay. a) The electric field distribution of the azimuthal field oscillates purely tangential to the ITO hole interface, resulting in no field enhancement at the ENZ interface. b) The transverse electric field for the radially polarized beam oscillates perpendicularly to the wall of the ENZ hole, yielding a field enhancement at all points along the interface.}
	\phantomsubcaption
	\label{fig:azi_hole}
	\phantomsubcaption
	\label{fig:rad_hole}
	\label{fig:hole_schem}
\end{figure*}

Despite the recent progress in characterizing the properties of the near-zero-permittivity regime \cite{niu2018epsilon,campione2015theory}, the experimental study of bare ENZ materials in the form of isolated nanostructures is largely unexplored so far. With the promising capabilities for significant nonlinear processes to take place on small footprints with ENZ materials \cite{reshef2019nonlinear,Abouelatta2023}, it is important to understand and characterize the linear optical response of ENZ nanostructures. In this work, we thus study ENZ nanostructures, i.e., nanoholes milled into an ENZ film, in the linear optical regime illuminated with sculpted light beams. Our ENZ material of choice is indium tin oxide (ITO), a low-loss TCO with its ENZ wavelength in the NIR regime. We choose the structure of the illumination such that it matches the cylindrical symmetry of the investigated nanostructures, while also allowing the realization of various interaction scenarios. We consider two orthogonal forms of structured beams \cite{Rubinsztein-Dunlop_2017} whose cylindrical symmetry matches that of the nanohole. It is shown that depending on the polarization of the incident beam, the transmission properties of the nanoholes contrast significantly.

With recent work exploring enhancement mechanisms between nanostructures and ENZ media \cite{issah2023epsilon,raad2024efficient,wang2020coupling,tirole2024nonlinear}, we are motivated, from a phenomenological perspective, to center our attention solely on the optical response of an individual ENZ nanostructure. Furthermore, we also showcase, for the first time, the potential of vectorially structured light in the interaction of electromagnetic waves with an ENZ material. A hole geometry is considered due to its straightforward fabrication and symmetry overlap with the incident field profiles we utilize. As we will see later, symmetry plays a significant role in either maximizing or minimizing the unique polarization-dependent effects observed. By understanding the transmission properties of these holes in the linear regime, one can further explore nonlinear schemes in the future with sure footing.

\section{Polarization- and Symmetry-Driven Field Enhancement}

The key property used throughout this work is centered around the boundary conditions for the normal component of an electric field oscillating at an interface: $\epsilon_{r,1}\mathbf{E}_{\perp,1} = \epsilon_{r,2}\mathbf{E}_{\perp,2}$, where the complex relative permittivity $\epsilon_{r,j} = \epsilon^{\prime}_{j} + i\epsilon^{\prime\prime}_{j}$, $j = 1,2$. Here the bulk properties of the two media are considered such that the surface charge density term $\sigma$ can be neglected. Solving for the electric field in medium $2$, where medium $1$ is vacuum and medium $2$ is the ENZ environment, the electric field takes the form: $\mathbf{E}_{\perp,2} = \frac{\epsilon_{r,1}}{\epsilon_{r,2}}\mathbf{E}_{\perp,1}$. As $\epsilon_{r,2} \rightarrow 0$, the normal component of the electric field in the ENZ environment diverges, resulting in a substantial field enhancement. 

ITO however possesses a non-negligible absorption coefficient, which solving for the system described above when $\epsilon_{r,1} = 1$ and $\epsilon^{\prime}_{2} \rightarrow 0$, results in $\mathbf{E}_{\perp,2} = -\frac{i}{\epsilon^{\prime\prime}_{2}}\mathbf{E}_{\perp,1}$. The enhanced field expected in the ITO therefore scales inversely solely with the imaginary component of the permittivity and is $-\frac{\pi}{2}$ out of phase with the incident field. Our measured ITO sample, which carries an imaginary component of the permittivity of $\epsilon^{\prime\prime}_{2} = 0.275$, consequently yields an enhancement in the electric field amplitude of $\sim\!3.6$. With the compelling potential ENZ materials offer owing to their field enhancements, much work has recently gone into characterizing and minimizing the losses in such systems \cite{li2019structural,anopchenko2023field}.

Oftentimes, the polarization distribution of the incident field, in conjunction with the geometry of the ENZ structure, is not configured for optimizing the field enhancement. Here we consider a geometry for the ENZ environment which overlaps symmetrically with the distribution of the electric field while examining sets of polarization states which either maximize or minimize the field enhancement. The structure which naturally facilitates a symmetric overlap is a cylindrical void in the ENZ medium. In order to obey the symmetry of the nanohole while probing the polarization-tailored responses, we individually excite each nanohole with either an azimuthally or radially polarized cylindrical vector beam \cite{Rubinsztein-Dunlop_2017,Zhan2009}. 

For a brief phenomenological interpretation, we only consider an interaction when the electric field distribution is perfectly centered with the nanohole, i.e., the optical axis of the beam being aligned with the symmetry axis of the hole. Under tight focusing, the azimuthally polarized beam retains qualitatively the focal electric field distribution in the transverse plane, where the electric field components oscillate tangential to the interior hole interface (Figure \ref{fig:azi_hole}). 

The tightly focused radially polarized beam in contrast leads to a more complex focal electric field distribution. It results in a 3-dimensional focal electric field with the transverse components maintaining a radial polarization distribution, overlaid with a strong field component oscillating in the longitudinal direction \cite{dorn2003sharper,Quabis2000,Youngworth2000}. Depending on the hole diameter, the longitudinal field can either interact strongly with the surface of the ENZ film, or overlap and interact with the glass substrate, leaving any interaction with the ENZ to the transverse field. In any case, for the radially polarized field, any field component which interacts with the ENZ interface will do so in a manner perpendicular to the interface itself, resulting in a field enhancement as indicated in Figure \ref{fig:rad_hole}.

\section{Experiment}

\begin{figure}[!h]
	\centering
	\includegraphics[width=.45\textwidth]{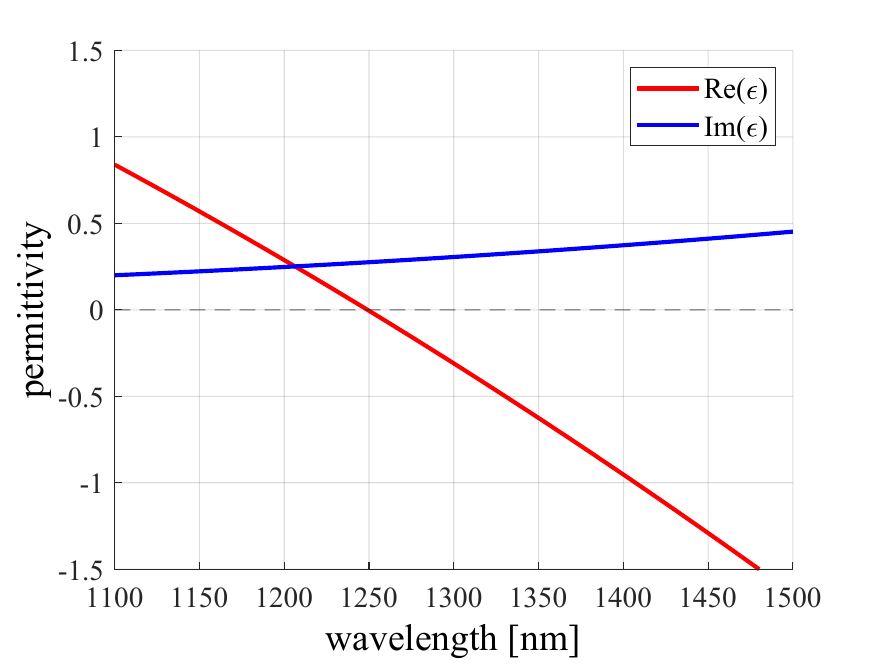}
	\caption{Experimental ellipsometry data taken for the unstructured ITO film. The real component of the permittivity crosses zero at $\lambda=\SI{1250}{nm}$.}
	\label{fig:itoperm}
\end{figure}

\begin{figure}[!h]
	\centering
	\includegraphics[width=.45\textwidth]{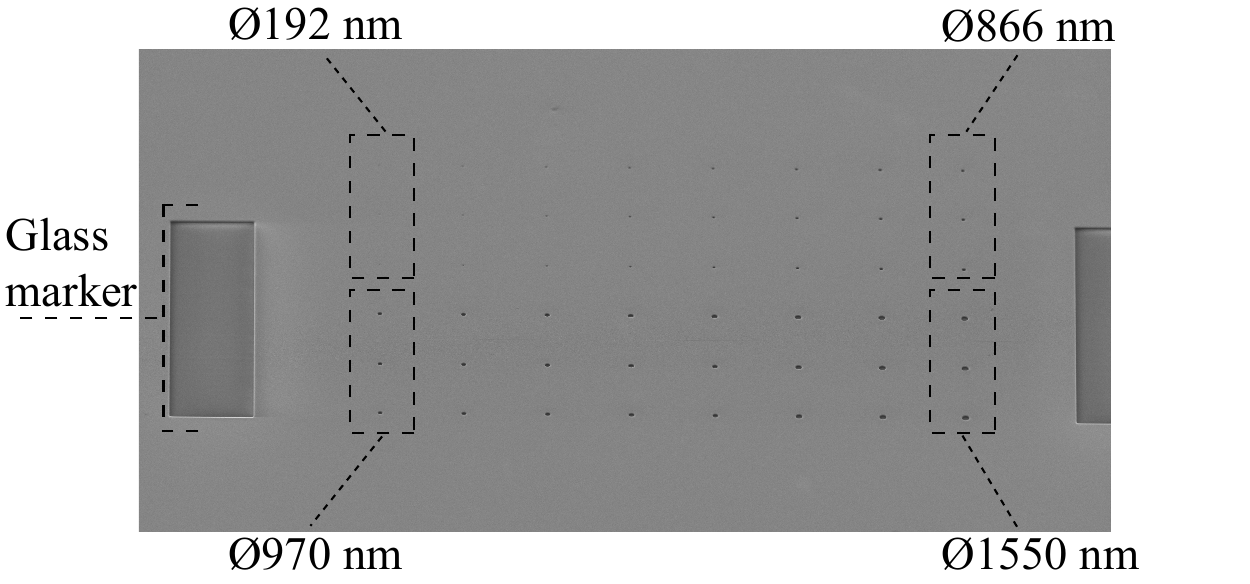}
	\caption{Scanning-electron micrograph of the milled nanoholes in the ITO film. For each hole diameter, two duplicate holes are milled.}
	\label{fig:nanoholes}
\end{figure}

Prior to any structuring, a $\SI{310}{nm}$ thick ITO film was sputtered on a $\SI{175}{\micro\meter}$ thick BK7 glass substrate \cite{pgo} and was placed in an ellipsometer to obtain the complex refractive index. The zero-permittivity value of the real part exists when the refractive index \textit{n} and the absorption coefficient \textit{k} are equivalent, which in this case occurs around $\lambda = \SI{1250}{nm}$, yielding $\widetilde{n} = 0.368 + i0.375$ (Figure \ref{fig:itoperm}). The film was then patterned with nanoholes via focused ion beam milling (Figure \ref{fig:nanoholes}). An array of holes with varying diameters were milled in the ITO film. In order to maintain the permittivity measured by the ellipsometer, it is imperative to minimize fabrication side effects from the focused ion beam. This is mainly achieved by using the correct ion species, which was found to be neon as opposed to the conventional use of gallium.

The nanoholes were designed to have diameters ranging from \SI{200}{nm} to \SI{1700}{nm} and were patterned using an optimized circular scanning strategy. All patterns are generated with "fib-o-mat" \cite{deinhart2021patterning}, carefully avoiding unintended electron and ion beam effects in the patterned areas before and after patterning. This included taking high-resolution scanning electron micrographs only on selected nanostructures whose position is noted such that they are not optically measured in experiment. Upon inspection it was found that most hole diameters were roughly \SI{30}{nm} smaller than their target diameter. The holes designed between \SI{1300}{nm} to \SI{1700}{nm} staggered in growth, resulting in a maximum diameter of $\sim$\SI{1550}{nm}. Despite this, the diameter range is still sufficient to allow us to observe the transmission for nanoholes both smaller and larger than the focal field. Three identical holes are milled per column in case a fabrication fault occurs for a given hole. The holes are spaced $\SI{20}{\micro\meter}$ apart from each other to ensure no near-field excitations take place. The measurement is done by performing a raster scan of the nanohole across the focal field with the use of the piezo stage. The end result is a scan image where each pixel corresponds to the relative power either reflected or transmitted for the beam with respect to the sample. The transmission images are normalized with respect to the glass substrate underneath the ITO film.

The optical setup begins with a super-continuum light source whose wavelength is selected using a liquid crystal wavelength filter. The output of the filter produces a beam with a wavelength of $\SI{1250}{nm}$ and a bandwidth of $\sim\!\SI{7}{nm}$. The beam passes through a single mode fiber to both guide the beam towards the main setup while also filtering the spatial mode of the incident field. Out of the fiber, the beam is launched into a tower configuration and guided to the top \cite{Banzer2010}. The field is first structured using a linear polarizer and a liquid-crystal-based q-plate, which allows one to either structure the beam to carry a radial or azimuthal polarization distribution \cite{marrucci2006optical}. A spatial filter is then placed shortly after to remove undesired higher-order spatial modes, refining the quality of the structured beam. 

\begin{figure}[t]
	\centering
	\includegraphics[width=.45\textwidth]{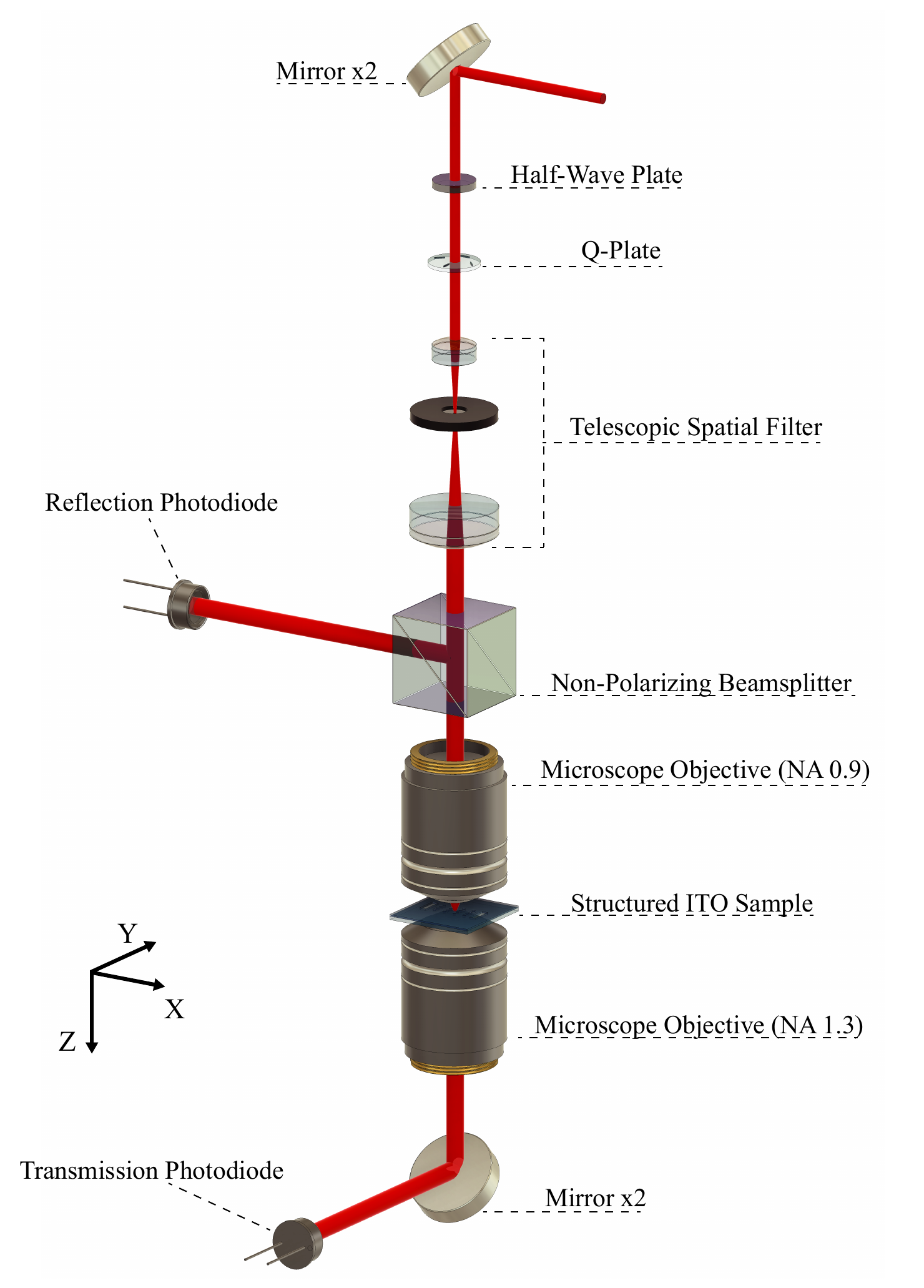}
	\caption{Experimental setup detailing the polarization structuring and tight focusing onto the structured ITO sample. The ITO sample is mounted to a 3D piezo stage (not depicted here) allowing for precise alignment of the nanoholes with respect to the focal field.}
	\label{fig:exp_setup}
\end{figure}
The filtered structured beam transmits through a non-polarizing beam splitter (NPBS) and is focused by a microscope objective with numerical aperture (NA) of 0.9. The field is tightly focused to a spot size on the order of the wavelength and interacts with the structured ITO sample which sits on a 3D piezo table. Light transmitted through the sample is collected with a 1.3 NA oil immersion objective, which due its confocal alignment, guides the collimated output towards a photodiode. Reflected light passes back through the focusing objective and reflects off the NPBS towards the reflection photodiode. While the end measurement data comes from the transmitted field, the measured reflected field is used for ensuring precise alignment between the focal field and the nanostructures.

The scan images allow us to realize various interaction scenarios and to relate the interaction strength with the field profile at the focus. The center of the scan image represents the transmitted power for the beam centered with the hole (referred to as on-axis position or illumination). When considering nanoholes in conductors such as silver, a waveguide interpretation can be assigned to model the transmission properties of these holes \cite{kindler2007waveguide}. Due to the enhancement effects of the field in the ENZ regime, such a model fails to hold fully in our case; therefore requiring numerical modeling by use of finite-difference-time-domain (FDTD) simulations. We mimic the experimental measurement procedure in our FDTD environment by raster scanning the nanohole on a glass substrate across a given structured focal field and collect the transmitted power with a field monitor for each position.

\section{Results}
\begin{figure*}
	\centering
	\includegraphics[width=.75\textwidth]{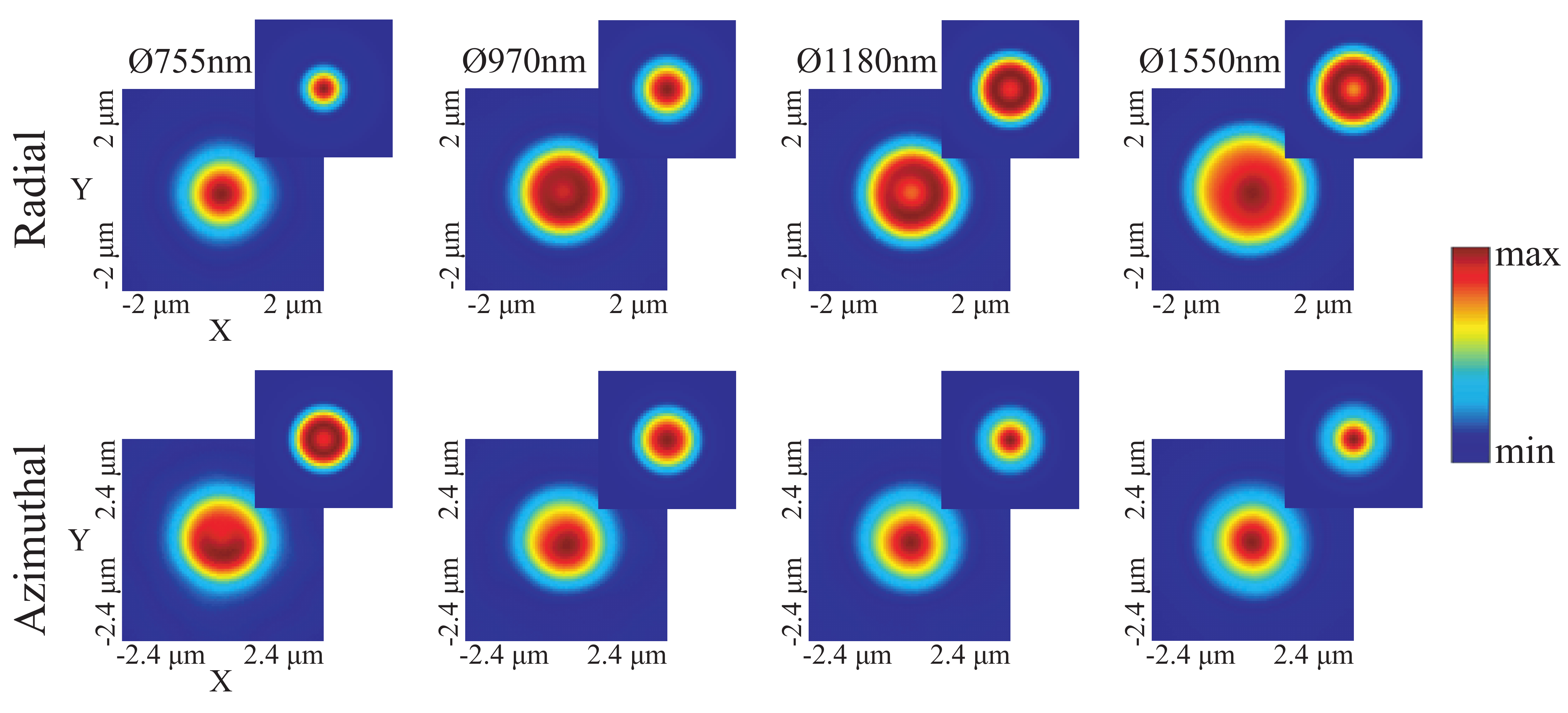}
	\caption{Experimental and simulated (insets) scan images spanning a selection of holes over the region of interest for an incident radially and azimuthally polarized beam. The central pixels represent purely on-axis transmission. All values are normalized to the transmission through the bare glass substrate with no ITO present.}
	\label{fig:scan_images}
\end{figure*}

\begin{figure}
	\centering
		\includegraphics[width=0.45\textwidth]{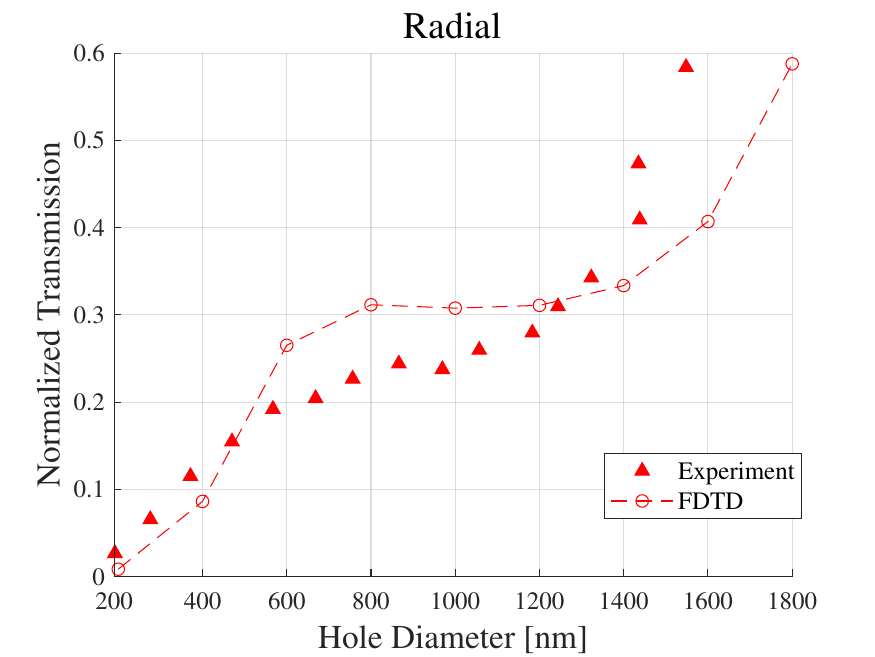}
		\caption{Transmission plot for the radially polarized beam focused on-axis with the center of the nanoholes. A plateau occurs for hole diameters where the strongest field overlap with the ITO (and therefore an enhancement) is expected.}
		\label{fig:trans_rad}
	\end{figure}

With the individual holes scanned in transmission, a qualitative and quantitative comparison between polarization distributions can be made for the transmission properties of these holes. For the radially polarized field, a striking feature in the transmission arises for hole diameters between $800 - \SI{1200}{nm}$. For on-axis illumination, the transmission stays almost constant in this region of hole diameters, indicated by both the scan images and the transmission plot (Figures \ref{fig:scan_images} \& \ref{fig:trans_rad}). The hole diameters where this effect occurs corresponds with the walls of the ITO hole overlapping spatially with the transverse field components of the focal field. This suggests that the suppressed transmission occurs when the transverse fields can strongly interact with the ITO. For the radially polarized field distribution, this indicates that the transverse field is oscillating normal to the ITO wall in all directions, generating a field enhancement in the ITO surrounding the hole. As the holes become large enough such that the transverse field components no-longer interact strongly with the ITO, the transmission begins to restore.

It is worth noting here that even for the largest hole diameter studied, the tightly focused beam still has some overlap with the ITO film for the chosen propagation parameters (NA, wavelength, etc.) for on-axis illumination. Thus, the recorded transmission still grows and does not reach yet a constant value for the largest holes tested (see Figure \ref{fig:trans_rad}). Furthermore, one sees that the unique transmission features for the experimental radial beam measurements occur earlier for holes smaller than what the FDTD predicts. Likewise, the transmission restores for the on-axis illumination earlier than what the FDTD shows. Still, an effective range where this interaction scenario takes place is shown well for the experimental data, however this range appears smaller compared to the FDTD results. Such discrepancies are left to local material changes induced by the FIB milling which were not modeled in this study.

\begin{figure}
	\centering
		\includegraphics[width=0.45\textwidth]{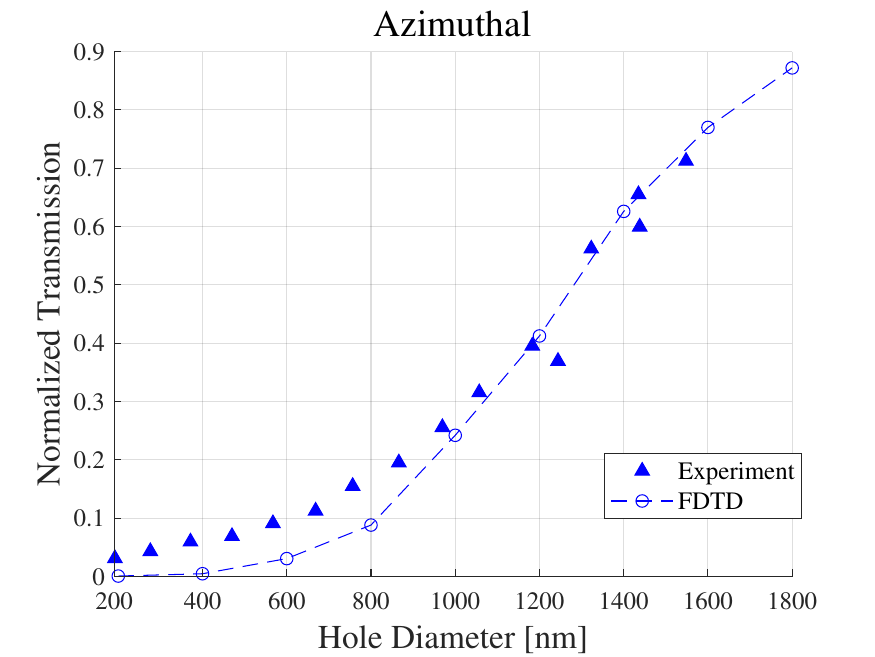}
		\caption{Transmission plot for the azimuthally polarized beam focused on-axis with the center of the nanoholes. The transmission experiences no such plateau and no field enhancements are expected in the ITO.}
		\label{fig:trans_azi}
\end{figure}

The results achieved for an azimuthally polarized input beam, however, contrast significantly with those for the radially polarized field (Figure \ref{fig:scan_images}). For most of the hole diameters, the on-axis transmission remains the most prominent in the performed scans. Only for smaller diameters is the on-axis transmission lower than for off-axis illumination (e.g., at $\SI{800}{nm}$). The on-axis transmission plotted with dependence on the hole diameter also indicates a similar behavior in the experimental and simulated data (Figure \ref{fig:trans_azi}). For the azimuthally polarized beam, the electric field in the interaction region is always parallel to the air-ITO interfaces for on-axis illumination, both with respect to the upper ITO surface and the ITO side walls. Hence, there exists no field component oscillating normal to the ITO interface, and consequently no field enhancement. 

\begin{figure}
	\centering
	\includegraphics[width=.5\textwidth]{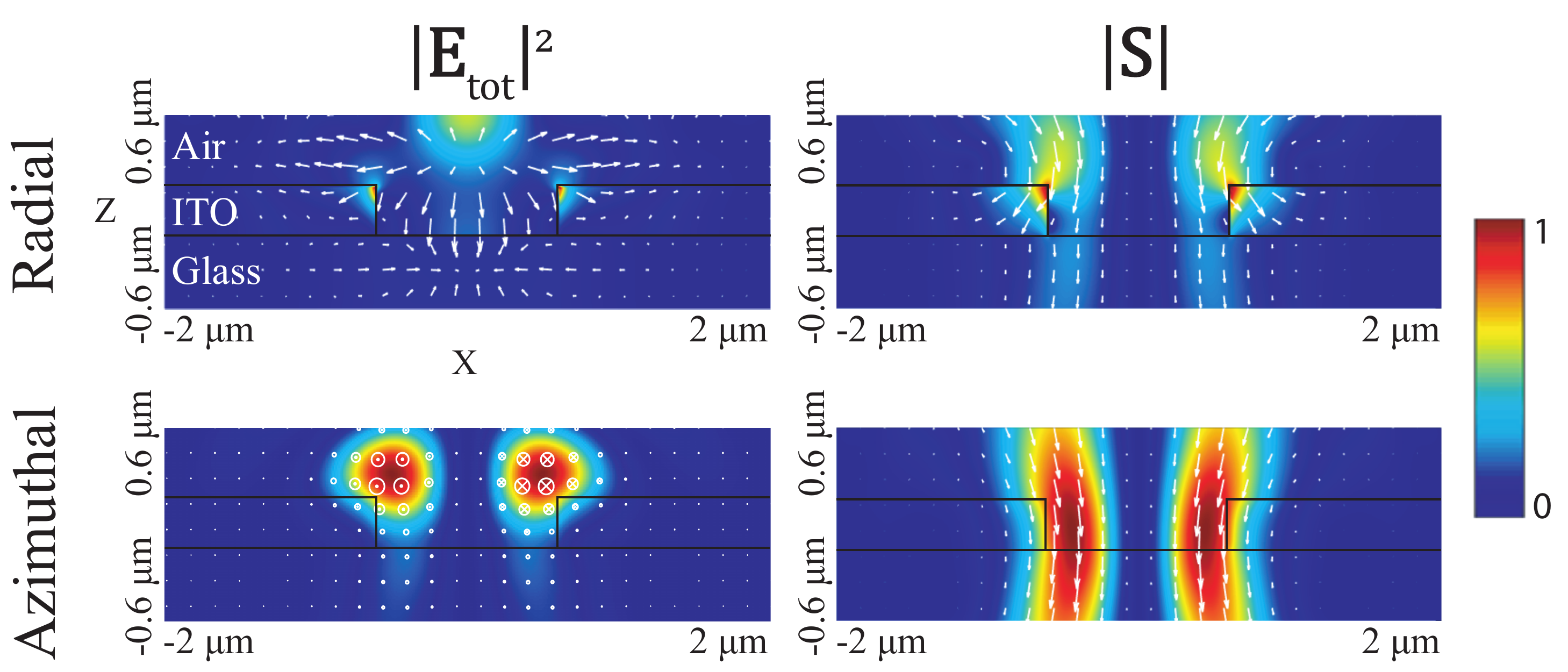}
	\caption{Cross-sectional view of the simulated time-averaged electric field intensity and Poynting field across a \diameter$\SI{1200}{nm}$ hole for both a tightly focused radially and azimuthally polarized beam. Electric field vectors and Poynting vectors are shown for a snapshot in time (arrow overlays).}
	\label{fig:E_S_fields_1200nm}
\end{figure}

To better gauge how the focal fields interact with the ENZ nanostructure, we performed a detailed numerical inspection of the electric fields and the energy flux (Poynting vector) in the interaction region. We show the corresponding cross-sections for a $\SI{1200}{nm}$ diameter hole illuminated by both a tightly focused radially and azimuthally polarized field (see Figure \ref{fig:E_S_fields_1200nm}). This hole diameter is considered due to the observed contrast in transmission between the two beams occurring here, and because for both beams the focal spot size is roughly equivalent to this diameter. Due to the strong lateral confinement of the longitudinal field for the focused radially polarized beam, the longitudinal component primarily interacts with the glass substrate, leaving any interaction with the ITO purely to the transverse field components oscillating perpendicular (normal) to the walls of the ITO. The electric field amplitude is enhanced at the wall and decays as the field extends into the ITO. 

The Poynting vectors, illustrated as white arrows in the respective plots in Figure \ref{fig:E_S_fields_1200nm}, features an enhanced transverse component in the corresponding region close to the side walls. Hence, the energy flowing through the hole and being detected in forward direction should be reduced significantly. This observation is consistent with the measured suppressed transmission for on-axis illumination. It should be noted here that the resulting drop in transmission for such an excitation scenario is not a direct consequence of the ENZ's field enhancement property. The transmission drop stems from a combination of factors, two of which are: poor confinement of the fields produced in the ITO layer due to the impedance mismatch from the surrounding larger refractive index media, and significant attenuation in the ITO due to its absorption. The drastic impedance mismatch between ENZ materials and dielectrics, however, has been shown to be an attractive feature for controlled out-coupling of quantum dot sources \cite{wu2019ultracompact}.

The electric field profile for the azimuthal beam is shown to be guided primarily through the hole, hardly influenced by the ITO. The maximum electric field amplitude remains to the free-space spatial mode, and no field enhancement is observed in the ITO. The Poynting vectors again support the observed behavior which was measured for this system. The energy dominantly flows along the optical axis, while the transverse vector components are a consequence of the significant nonparaxiality due to the high NA objective. An increased transverse energy flow resulting from the light-ITO interaction is not observed. 

\begin{figure}[h]
	\centering
	\includegraphics[width=.5\textwidth]{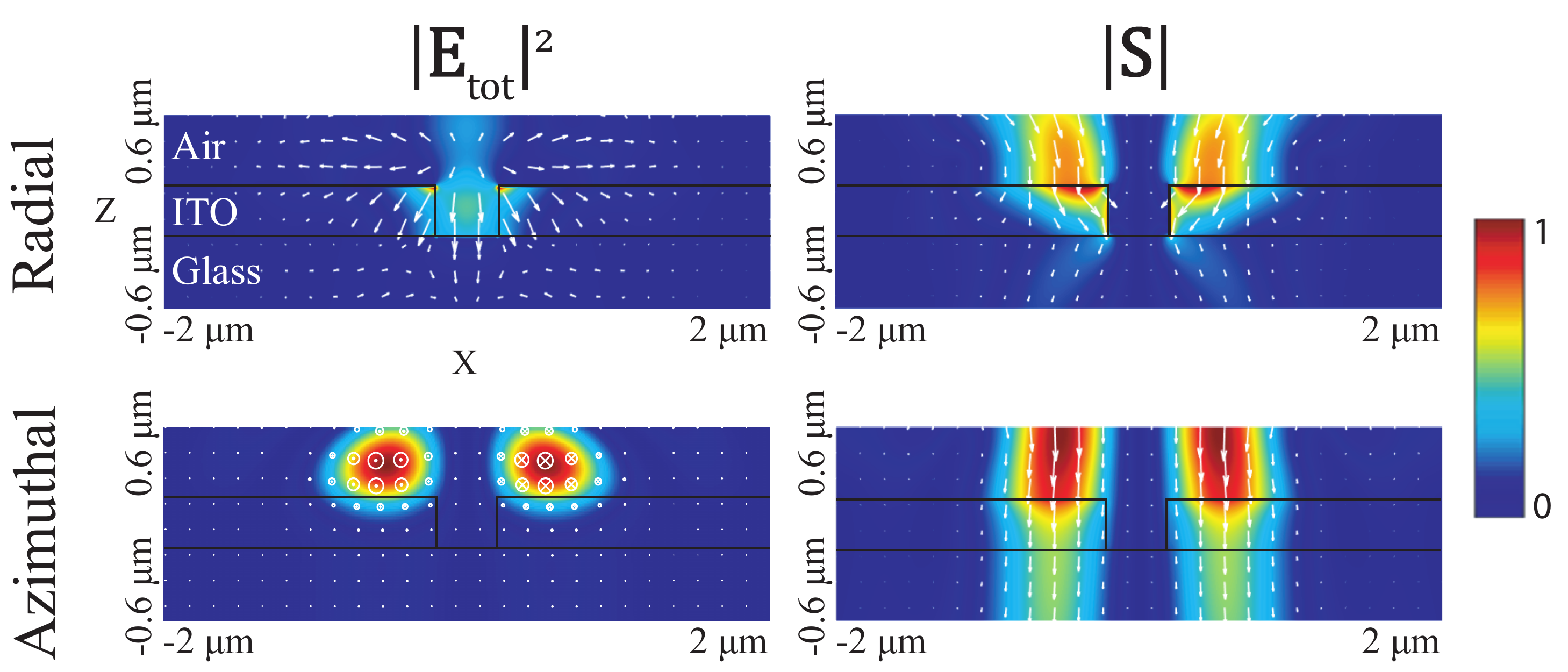}
	\caption{Cross-sectional view of the simulated time-averaged electric field intensity and Poynting field across a \diameter$\SI{400}{nm}$ hole for both a tightly focused radially and azimuthally polarized beam.}
	\label{fig:E_S_fields_400nm}
\end{figure}

The field dynamics shown in the FDTD results indicate agreement with the experimentally observed transmission properties of the nanoholes. When considering hole diameters much smaller than the spot size of the focal field ($<\SI{800}{nm}$), the interaction with the radially polarized field is significantly affected (Figure \ref{fig:E_S_fields_400nm}). In this scenario, the longitudinal component, which interacts with the upper surface of the ITO layer, contributes to the field enhancement as well. These enhanced field components modify the total Poynting vector distribution in the ITO layer, resulting in a net energy flow which converges towards the optical axis, as opposed to larger holes whose Poynting vectors divert from the optical axis. 

\begin{figure}[h]
	\centering
	\includegraphics[width=.5\textwidth]{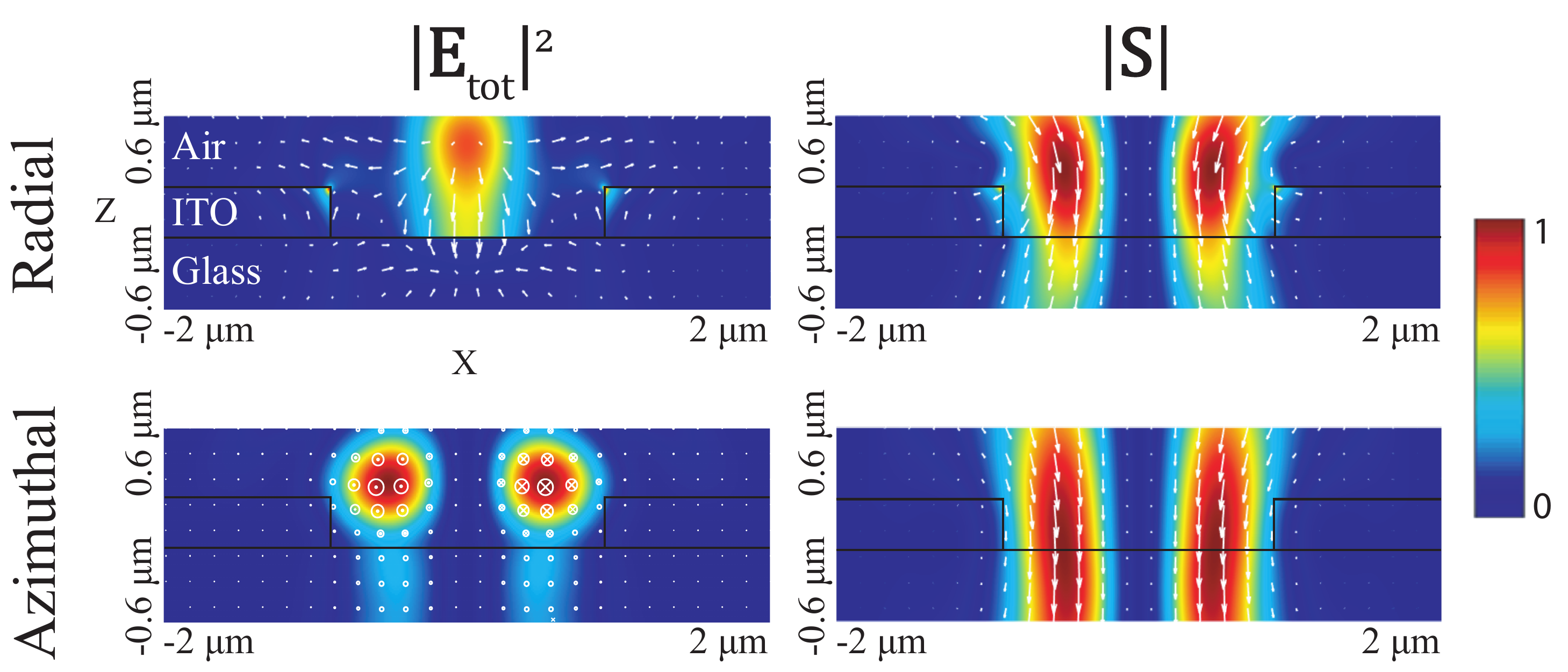}
	\caption{Cross-sectional view of the simulated time-averaged electric field intensity and Poynting field across a \diameter$\SI{1800}{nm}$ hole for both a tightly focused radially and azimuthally polarized beam.}
	\label{fig:E_S_fields_1800nm}
\end{figure}

For hole diameters much larger than the effective interaction area (Figure \ref{fig:E_S_fields_1800nm}), the radially polarized field weakly interacts with the walls of the ITO nanohole, allowing the beam to transmit through largely unaffected and therefore restoring the transmission. It is only once the hole diameter is large enough to neglect the interaction between the strong longitudinal field and ITO, while still being small enough to facilitate the interactions with the transverse field components, that the energy is partially redirected transversely into the ITO film and the net transmission drops.

\section{Conclusion}
The unique consequences of the boundary conditions for an ENZ medium indeed give rise to significant contrasting effects depending on the structure of the medium and the polarization distribution of the incoming excitation field. For nanoholes in an ENZ film, focal fields with cylindrical symmetry can either lead to strong boundary-driven field enhancement effects or completely suppress them depending on the polarization. This behavior was experimentally shown and numerically verified here for tightly focused radially and azimuthally polarized field distributions incident on such holes. Already in the linear optical regime, the contrasting transmission properties enabled by the ENZ nanostructure and focal field provide a suitable environment for polarization-based optical switching schemes. Other structures, such as co-axial nanoholes, may also serve useful by allowing the longitudinal component of a focused radially polarized beam to fully interact with the ITO surface, while still allowing an interaction between the transverse field and the ITO's side walls. Furthermore, our results serve as a basis for exploring next steps in the nonlinear regime, where significant nonlinear refractive index changes occur which are facilitated by the field enhancement mechanisms discussed above. Under the construction used here, field-enhancement-driven refractive index changes could be achieved with substantially less incident power due to taking advantage of the polarization-dependent field enhancement by the ENZ structure, resulting in efficient polarization-enabled all-optical switches and routers.

\section*{Acknowledgments}
The financial support by the Austrian Federal Ministry of Labour and Economy, the National Foundation for Research, Technology and Development and the Christian Doppler Research Association is gratefully acknowledged. Victor Deinhart and Katja H{\"o}flich acknowledge the financial support from the Leibniz Foundation (project ENGRAVE Nr. K335/2020). The He ion beam patterning was performed in the Corelab Correlative Microscopy and Spectroscopy at the Helmholtz-Zentrum Berlin and within the framework of the EU COST action CA 19140.

\section*{Conflict of Interest}
The authors declare no conflicts of interest.

\section*{Data Availability Statement}
Data regarding the results presented in this article are not publicly available at this time but may be obtained from the authors upon reasonable request.

\bibliography{bibliography}

\end{document}


\maketitle
\vspace{-30pt}
\section{Additional simulations}

\begin{figure}[!h]
	\centering
	\begin{subfigure}{0.5\textwidth}
	\includegraphics[width=\textwidth]{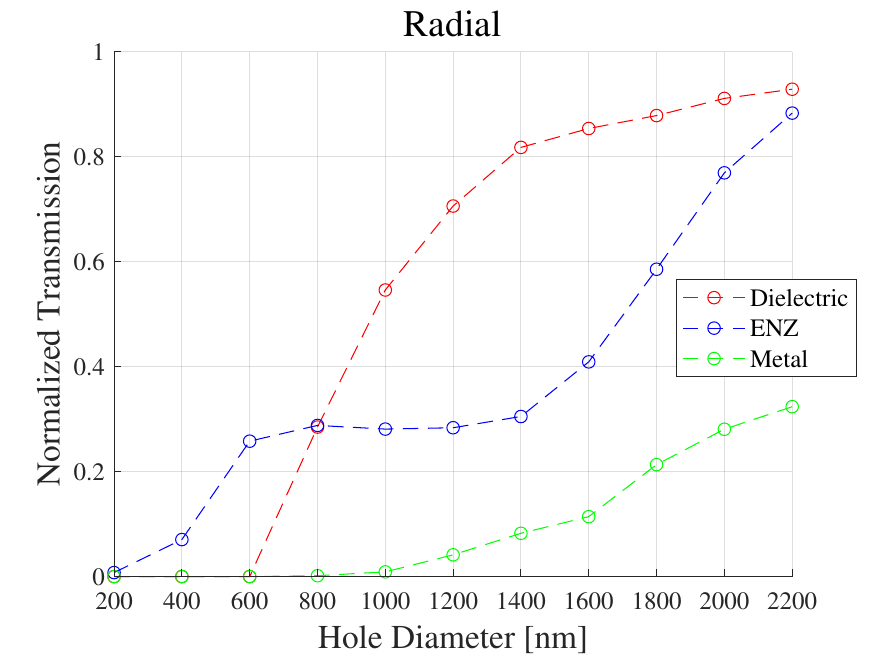}
	\end{subfigure}%
	\begin{subfigure}{0.5\textwidth}
	\includegraphics[width=\textwidth]{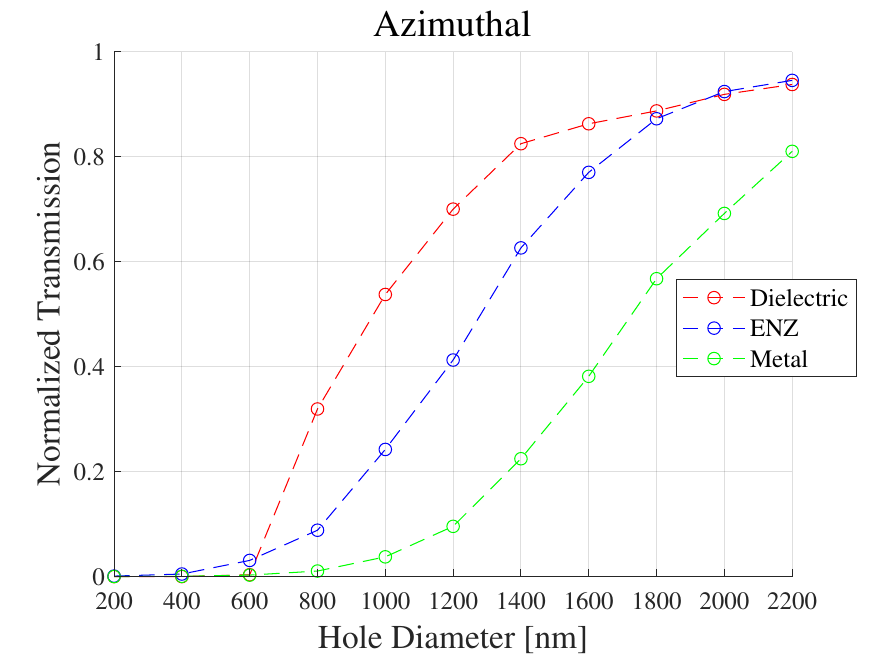}
	\end{subfigure}
	\caption{FDTD results for the on-axis spectral-dependent transmission of ITO nanoholes over varying hole diameters for both radial and azimuthal beams.}
	\label{fig:spectral_response}
\end{figure}

Figure \ref{fig:spectral_response} shows FDTD results for the transmission of the ITO nanoholes across hole diameter for varying wavelengths. Three wavelengths were chosen from the ellipsometry data to entertain where the ITO behaves either as a dielectric, an ENZ medium, or a metal. The plots show that the transmission behavior for the radially polarized beam is indeed unique to the ENZ response of the ITO. At $\lambda = \SI{650}{nm}$, the complex refractive index of the ITO is $\widetilde{n} = 1.67 + i0.287$ and we treat the ITO as a lossy dielectric. The ENZ wavelength is $\lambda = \SI{1250}{nm}$ and the complex refractive index is $\widetilde{n} = 0.368 + i0.375$. While the absorption is still low for the longer wavelengths, we still treat the ITO as a metallic material at $\lambda = \SI{1650}{nm}$, primarily due to $\Re({\epsilon})$ being negative as $\Im({\epsilon})$ grows. With that, the complex refractive index at $\lambda = \SI{1650}{nm}$ is $\widetilde{n} = 0.177 + i1.67$.

\begin{figure}[!h]
	\centering
		\includegraphics[width=0.85\textwidth]{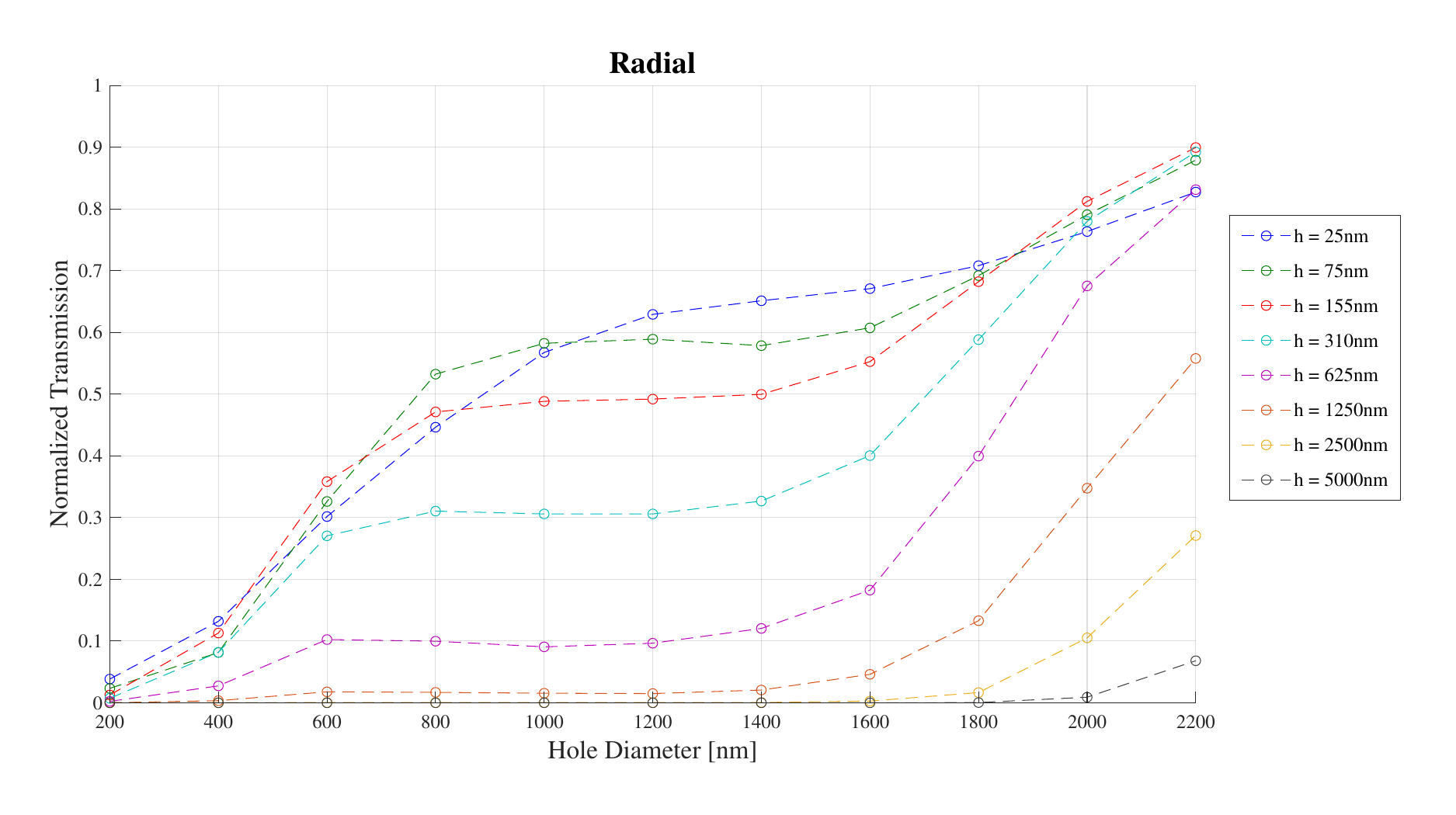}
	\caption{Transmission for a radially polarized beam with respect to hole diameter at the ENZ wavelength for varying ITO thicknesses.}
	\label{fig:radial_thickness_sweep}
\end{figure}

\begin{figure}[!h]
	\centering
 \vspace{-5pt}
	\includegraphics[width=0.85\textwidth]{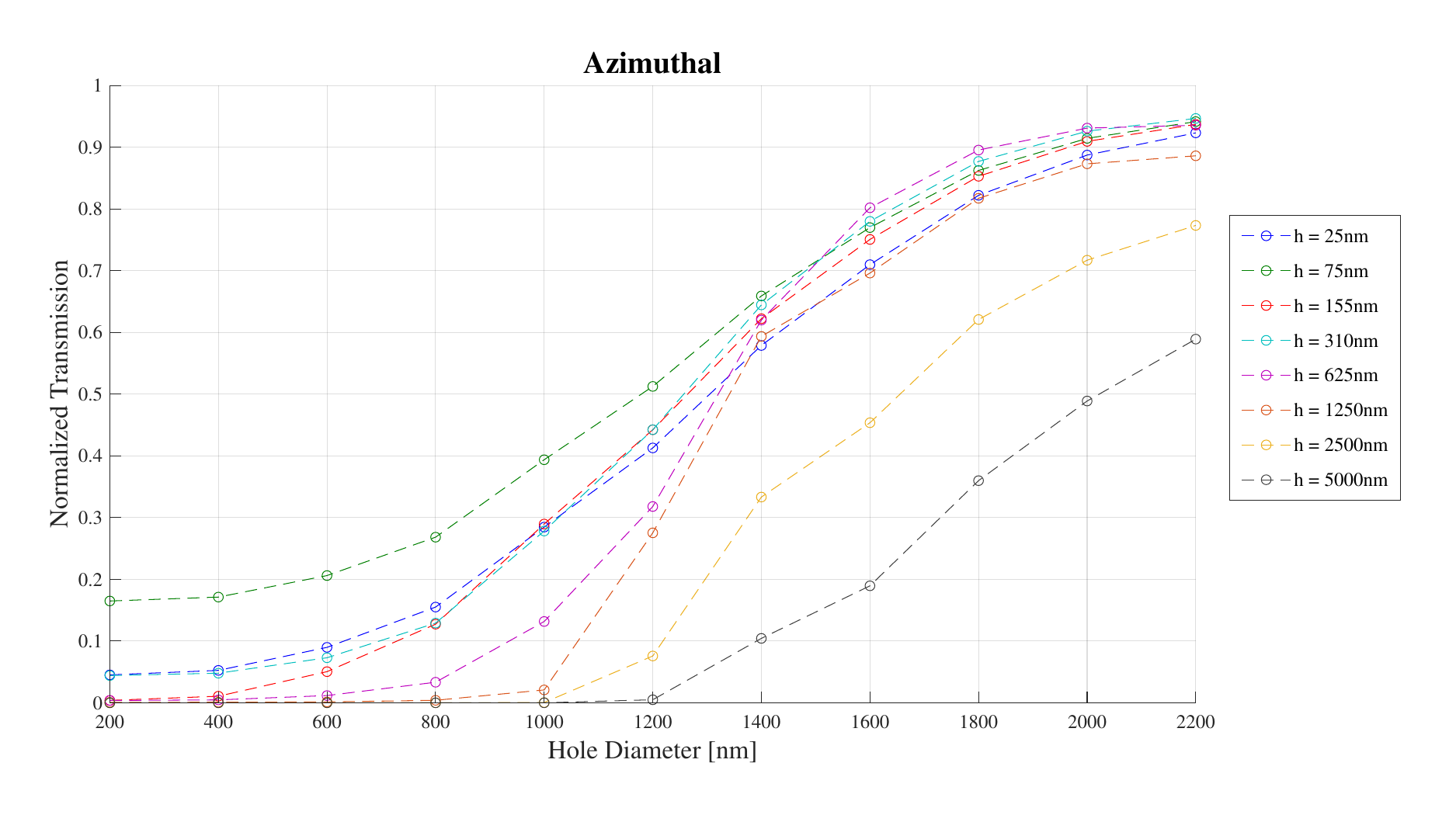}
	\caption{Transmission for an azimuthally polarized beam with respect to hole diameter at the ENZ wavelength for varying ITO thicknesses.}
	\label{fig:azimuthal_thickness_sweep}
\end{figure}

Figures \ref{fig:radial_thickness_sweep} and \ref{fig:azimuthal_thickness_sweep} show a thickness (denoted by $\textrm{h}$) sweep of the ITO which was performed in an FDTD environment at the ENZ wavelength. Here we see for both cases, and especially the radial beam, as the ITO thickness increases, the transmission becomes attenuated by the loss mechanisms of the material. For the radially polarized beam at $\textrm{h} = \SI{25}{nm}$, the ITO becomes too thin to facilitate the polarization-driven effects under study. Furthermore, additional mechanisms such as Berreman and ENZ modes may become present \cite{vassant2012berreman,campione2015theory}, modifying the transmission properties further. The primary transmission characteristic remains for most thicknesses with an incident azimuthally polarized focal field. The transmission however becomes strongly attenuated for thicknesses greater than $\textrm{h} = \SI{625}{nm}$.

\section{Sample fabrication}
\begin{figure}[!h]
	\centering
	\includegraphics[width=0.95\textwidth]{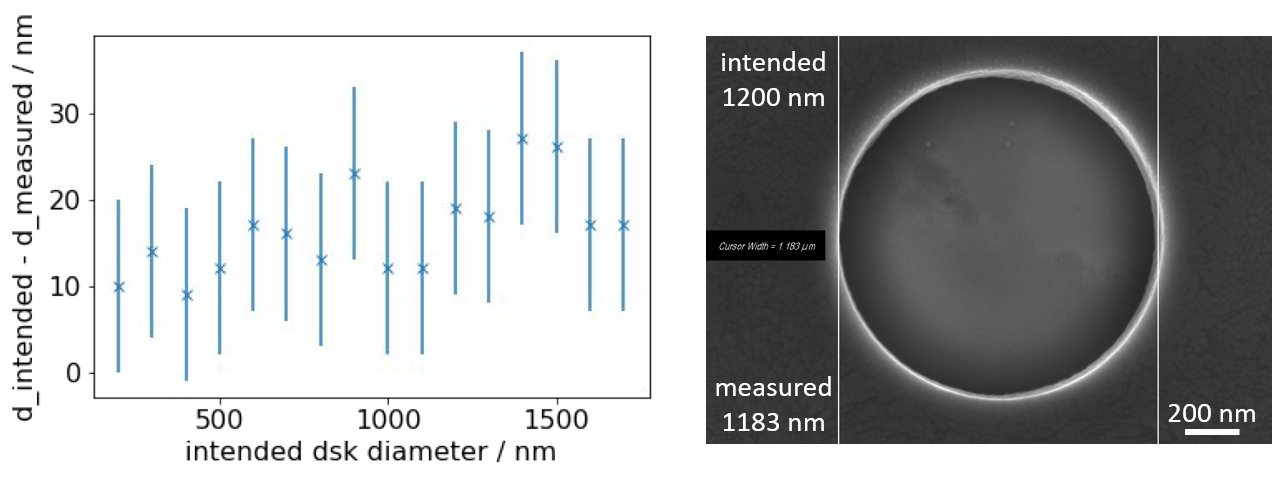}
	\caption{Obtained deviations over intended hole diameters and example of a high-resolution scanning electron micrograph used for the measurement.}
	\label{fig:meas}
\end{figure}

Ne-ion patterning was performed using a Zeiss Orion Nanofab microscope.
At an acceleration voltage of 26 kV, a gas pressure of 2\,x\,10$^{-6}$\,Torr was applied, resulting in an ion beam current of 3.9 pA on the sample surface.
The beam path was programmed to move  along a contour-parallel path that spiraled in and out of the circular pattern.  
The distance between two successive beam positions at which the beam dwelled for 0.1 \textmu s was 1\,nm. 
The optimal dose of 2.25\,nC/\textmu m$^2$ was calibrated based on a systematic dose variation, followed by mapping the obtained topography using atomic force microscopy.
The size of the patterned circular area had to be corrected for the expected increase in size due to the finite width of the material removal areas caused by the beam.
This led to a slight overcorrection, as can be seen in Figure~\ref{fig:meas}, where the hole diameters are systematically too small by about 10\,-\,30\,nm.

\bibliography{supp_bibliography}